# Schrödinger electrons interacting with optical gratings: quantum mechanical study of the inverse Smith–Purcell effect

Nahid Talebi

*Max Planck Institute for Solid State Research, Stuttgart Center for Electron Microscopy, Heisenbergstr. 1, D-70569 Stuttgart, Germany*
E-Mail: n.talebi@fkf.mpg.de

**Abstract-** Slow swift electrons with low self-inertia interact differently with matter and light in comparison with their relativistic counterparts: they are easily recoiled, reflected, and also diffracted form optical gratings and nanostructures. As a consequence, they can be also better manipulated into the desired shape. For example, they get bunched quite fast in interaction with acceleration gratings in presence of an external electromagnetic radiation, a phenomenon which can be desirable in development of superradiant coherent light sources. Here, I examine the spatiotemporal behavior of pulsed electron wave packets at low energies interacting with pulsed light and optical gratings, using a quantum-mechanical self-consistent numerical toolbox which is introduced here. It will be shown that electron pulses are accelerated very fast in interaction with the near-field of the grating, demanding that a synchronicity condition is met. To prevent the electrons to be transversely deflected from the grating a symmetric double-grating configuration is necessary. It is found that even in this configuration, diffraction due to the interaction of the electron with the standing-wave light inside the gap between the gratings, is a source of defocusing. Moreover, the longitudinal broadening of the electron pulse directly affects the final shape of the electron wave packet due to the occurrence of multiple electron-photon scatterings. These investigations pave the way towards the design of more efficient electron-driven photon sources and accelerators.

1.   **Introduction**

Electron-photon interactions have been the subject of intense studies. Several spontaneous and stimulated mechanisms of radiation from electrons have been hitherto detected; i.e. Larmor radiation [1], Cherenkov radiation [2], bremsstrahlung [3], transition radiation [2], Smith–Purcell effect [4], and stimulated Compton–Raman scattering [5], which have found applications in particle detectors [6], efficient electron-driven photon sources [7-9], and spectroscopy techniques [10]. Moreover, a combined system of laser beams and electron wave packet in the presence of matter which mediates the electron–photon interaction, causes the electrons to also absorb photons. This has been indeed perfectly demonstrated in the recent field of photon-induced near-field electron microscopy (PINEM) [11-14]. In addition to PINEM, absorption of photons by electrons leads to energy gain, a phenomenon which is incorporated and optimized in design of linear [15-17] and dielectric laser accelerators (DLAs) [16, 18, 19], and also streak cameras [20, 21].

Theory of Compton scattering clearly indicates that an inelastic interaction of electrons with photons cannot occur with a concomitant propagation of electrons and plane wave light in free space [11]. In order to boost the interaction and facilitate the necessary momentum and energy transfer, the near-field distribution of the optical modes is usually considered as the interaction medium [10]. In particular, the evanescent field in the vicinity of discontinuities such as an interface [22-24], plasmonic nanostructures [25-35], photonic crystals [36, 37], and gratings [38, 39] has been so far considered. Interestingly, a classical treatment of the electron–photon interaction demonstrates the necessity of the momentum conservation criterion to be met as $\omega_{ph} = \vec{k}_{ph} \cdot \vec{V}_e$ ($\omega_{ph}$ is the angular momentum of the photon, $\vec{k}_{ph}$ is the wave vector of the photon, $\vec{V}_e$ is the velocity of the electron) which agrees perfectly well with the first principles of quantum mechanics [10]. When a grating is incorporated as the medium of interaction, the dispersion of the electron-induced polarization is mapped into the optical cone as $\omega_{ph} = \left(\vec{k}_{ph} + \hat{\alpha} 2m\pi / \Lambda \right) \cdot \vec{V}_e$, according to the Floquet's theorem. Here $\hat{\alpha}$ is a unitary vector along the grating axis, $\Lambda$ is the period of the grating, and $m$ is an integer denoting the diffraction order. Indeed this mapping is the reason behind the Smith–Purcell radiation into the far-field. When an external laser field is also applied to the grating, an electron traveling adjacent to the grating may gain energy by absorbing photons, a phenomenon referred to as inverse Smith–Purcell effect [40].

Although classical theory can be routinely applied to investigate the photon emission process in the absence of an external radiation, it cannot be applied to photon-induced emissions, as for PINEM [41-44]. Especially, the probability of the electron to emit or absorb several quanta of photons in a multiple scattering process can only be treated quantum mechanically [44], which underlines the progress in the description of PINEM experiments through the computation of quantum mechanical propagators by considering either the Lippmann–Schwinger equation [42, 45], or making use of the scattering theory [11]. In this regard, quantum mechanics should be also applied to study the inverse Smith–Purcell effect. Although Lorentz force can be used to understand the linear acceleration in interaction of electrons with the synchronized mode of a grating, it cannot describe the interferences and also not the bunching which happens due to multiple scattering and nonlinear processes within one single-electron wave packet. Therefore, classical mechanics cannot describe the chirping of single electron wave packets due to such interactions.

The reason that such quantum mechanical approaches have been not yet employed to study the inverse Smith–Purcell effect is mostly due to the very short de Broglie wavelength of electrons at high energies compared to optical wavelengths and dimensions of employed devices, and partly due to the lack of an equivalent and stable particle-in-cell numerical approach in quantum mechanics. In fact, self-consistent Maxwell–Lorentz simulations and particle-in-cell numerical approaches are routinely applied to study the interaction of electrons with electromagnetic waves in electron-driven photon sources and accelerators [46-48]. Here, a step towards the development of a self-consistent Maxwell–Schrödinger numerical toolbox is reported and exploited to investigate the inverse Smith–Purcell effect. The advantage of such a numerical approach is manifold: first, it can be used to understand the quantum effects in interaction of electron wave packets with matter and external radiation; second, it can serve as a benchmark for analytical treatments such as adiabatic assumptions [49] and the Wolkow solution [50]; third, it can be used for understanding the dynamics during the interaction, the very same reason for which ultrafast experiments are routinely exploited.

Here, using the proposed Maxwell–Schrödinger numerical toolbox, the interaction of a single electron wave packet with light and gratings will be studied. It will be shown that a chirped grating would be necessary to avoid fast dephasing of electrons. Moreover, the effect of the electron pulse broadening on the shape of the electron wave packet after its interaction with the grating will be discussed. It will be also demonstrated that a symmetric grating configuration, though prohibiting the electron from being deflected away from the grating [51, 52], imposes another purely quantum mechanical effect on especially slow single electron wave packets: in fact the electron waves will be diffracted symmetrically in the transverse direction, via a two-photon processes, because of the Kapitza–Dirac effect [53-55].

As it will be discussed in Sec. II, a self-consistent numerical approach should be time-dependent in order to account for pulsed optical and matter waves and nonlinear effects. In Sec. III, the interaction of a pulsed electron with a silicon grating and pulsed laser illumination will be numerically investigated. Moreover, in section IV, the effect of the broadening of the electron pulse on the mechanism of the acceleration will be discussed. In section V, the conclusion and outlook for further investigations will be described.

## 2. Method

Time-dependent calculations have several advantages over the time-harmonic scattering theory, both in understanding the physical phenomena and interpretation, and also in simulating the nonlinear interactions triggered by pulsed wave functions. Both in molecular dynamics calculations and in the field of electromagnetics, there exists a plethora of such methods, from the Fourier-method [56] to finite-difference time-domain (FDTD) approach [57-59], and most recently the discontinuous Galerkin time-domain method [60]. Here, by combining the two molecular dynamics and electromagnetics simulation domains, a self-consistent numerical toolbox is developed, in order to investigate the evolution of matter wave-packets interacting with electromagnetic waves and nanostructures.

The numerical simulation domain is divided into two individual sub-domains, namely the Schrödinger domain and the Maxwell domain. The Schrödinger equation in the presence of the electromagnetic interaction is given by:

$$\left( -\frac{\hbar^2}{2m_0}\nabla^2 - \frac{i\hbar e}{m_0}\vec{A}(\vec{r},t)\cdot\vec{\nabla} + \left|\vec{A}(\vec{r},t)\right|^2 \right.$$
$$\left. -e\varphi(\vec{r},t) \right)\psi(\vec{r},t) = i\hbar\frac{\partial\psi(\vec{r},t)}{\partial t} \tag{1}$$

where the Coulomb gauge is applied. Here, $m_0$ is the free electron mass, $\psi(\vec{r},t)$ is the electron wave function, $\vec{r}$ is the displacement vector, $t$ is time, $e$ is the electron charge, $\vec{A}(\vec{r},t)$ and $\varphi(\vec{r},t)$ are the electromagnetic vector and scalar potentials, respectively. In order to numerically solve eq. (1), a pseudospectral method which is known as the Fourier Method is used [56]. In such an approach, the spatial differentiations are performed iteratively in the spatial-frequency domains as $\partial\psi(\vec{r},t)/\partial\alpha = \sum_n ik_{\alpha,n}\tilde{\psi}_n^\alpha(t)\exp(ik_{\alpha,n}\alpha)$, where $\alpha \in (x, y, z)$, $k_\alpha = 2n\pi/L_\alpha$, $L_\alpha$ is the size of the simulation domain in the direction $\alpha$, and $\tilde{\psi}_n^\alpha(t) = \int d\alpha\, \psi(\vec{r},t)\exp(-ik_\alpha\alpha)$, where by exploiting the orthogonality of the Fourier functions with equidistant sampling points, the integration can be approximated by a summation [61]. Time propagator however, is approximated by the second-order-differencing scheme [61]. Clearly this approach imposes a periodic boundary condition on the simulation domain. To avoid this, an absorbing boundary condition (ABC) is introduced which mimics a radiation boundary condition [62]. It has been shown in ref. 62, that an effective ABC can be formulated by an amplitude reduction in the form of $\psi(x,y;t+dt) = (1-\gamma\delta t)\psi(x,y;t)$, where $\gamma = U_0\left(1/\cosh^2(\beta x) + 1/\cosh^2(\beta(x-L_x))\right) + U_0\left(1/\cosh^2(\beta y) + 1/\cosh^2(\beta(y-L_y))\right)$, where $L_x$ and $L_y$ are the lengths of the simulation domain in $x$ and $y$ directions, respectively. The numerical investigations shown here that the choice of $\beta = 0.2\delta x$ and $U_0 = 0.02/e$ leads to an appropriate ABC, where $\delta x$ is the discretization length. Moreover, throughout the paper, all the simulations are two-dimensional. As the polarization of the incident optical wave is along $x$-direction, only $TE_z$ optical modes are excited, hence two-dimensional simulation is a justified simplification. Moreover, considering the only nonzero components of the electromagnetic field ($E_x$, $E_y$, and $H_z$), the momentum transfer from the optical modes to the electron will be only within the $x$-$y$ plane, hence the two-dimensional simulation for the electron wave function also suffices. In addition, throughout the whole paper, $\psi(\vec{r},t)$ is the wave function of a single-electron wave packet. Generalization of the present method to the case of multiple electron wave functions can be realized by modifying the scalar potential $\varphi(\vec{r},t)$ to include the interaction between different electron wave functions, which will be the topic of another relevant contribution.

The electromagnetic potentials in eq. (1) however, are unknown themselves and should be solved using the Maxwell equations. Moreover, when the electrons self-field is not negligible, the current and charge density distributions are themselves sources of electromagnetic fields. In this regard, eq. (1) and the Maxwell equations form a combined system of equations, which can be approached numerically in a self-consistent way (see figure 1). Time-dependent calculations for the Maxwell domain is performed by considering the FDTD method, using the previously developed numerical codes, which can include anisotropic and nonlinear domains, as well as materials with magneto-electric effect [58, 63]. Obviously, the connections between these two simulation domains are the current distribution function and the electromagnetic potentials, which themselves are calculated easily using the electron wave function and the electromagnetic fields (see figure 1). In addition to the electron current source, external electromagnetic radiations are also considered as photon sources denoted by $\vec{J}_{ph}$, or in an additive way, by using Huygens' principle [43]. The permittivity of different materials is modelled by the Drude function in addition to two critical point functions [57, 58].

Indeed these are electromagnetic potentials, but not the field components, which are to be included in the Hamiltonian of the interaction. To compute electromagnetic potentials from the field components, an inverse method should be employed. Considering the Coulomb gauge, a Laplace's equation for the scalar potential is obtained as $\nabla^2 \varphi(\vec{r},t) = -\vec{\nabla} \cdot \vec{E}(\vec{r},t)$. To obtain $\varphi(\vec{r},t)$, a finite-differentiation algorithm using Dufort-Frankel approach has been considered which is unconditionally stable. After obtaining $\varphi(\vec{r},t)$, the vector potential $\vec{A}(\vec{r},t)$ is obtained using the equation $\partial \vec{A}(\vec{r},t)/\partial t = -\vec{\nabla}\varphi(\vec{r},t) - \vec{E}(\vec{r},t)$.

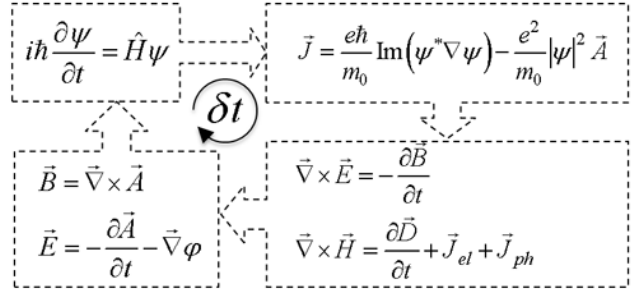

**Figure 1**. The schematic of a self-consistent algorithm, developed to calculate the electron wave functions interacting with potentials.

3. **Results and discussion**

As experimentally demonstrated by Furuya and coworkers [40], an electron propagating parallel to a grating can be accelerated once the grating is illuminated with an external laser radiation if the synchronicity condition is met. These observations were the driving force behind the development of a new class of linear accelerators (linac), with acceleration gradients as high as 100 MeV m$^{-1}$ [64]. As it has been recently demonstrated, reaching even higher acceleration gradients would be possible by incorporating dielectric gratings [19, 65]. Dielectrics in comparison with metals offer a higher damage threshold at optical frequencies. In addition, ultrafast lasers operating at optical frequencies are so stable that even sub-cycle carrier-envelope phase effects can be investigated. Because of these facts, dielectric laser acceleration is emerging as the next-generation linac.

A parallel field of study in wave science is ultrafast electron microscopy using low-energy photoemission electron sources [66-76]. Pulsed electrons from photoemission sources are highly coherent and can be directly exploited in ultrafast imaging and diffraction, and also in spectroscopy [20, 26, 75, 77-81]. As will be shown here, low-energy pulsed electrons are also better accelerated and shaped, because they are more sensitive to the electromagnetic radiation and their interaction with the laser excitation will take place in a coherent way. Hereafter, the interaction of low-energy electron pulses with a dielectric grating will be investigated.

A two-dimensional silicon grating composed of silicon nanorods, each with the width of 16 nm and a height of 20 nm, as shown in figure 2(a), is considered here as the acceleration medium. The grating is illuminated with a linearly polarized laser pulse at the carrier wavelength $\lambda_0 = 830\,\text{nm}$, peak electric field amplitude of 0.5 GV/m, and with a broadening of 80 fs. An electron pulse at the initial carrier energy of 408 eV ($V_e = 0.04c$, where $c$ is the speed of light) traveling parallel to the grating axis and at a distance of 4 nm from the grating is considered. The initial distribution of the electron pulse is assumed to be a two-dimensional Gaussian function as:

$$\psi(\vec{r}, t=0) = \exp(ik_x^e x) \left( \frac{1}{\sqrt{2\pi} W_t} \exp\left(-\frac{1}{2} \frac{(x-x_0)^2}{W_t}\right) \right)^{\frac{1}{2}} \times \left( \frac{1}{\sqrt{2\pi} W_L} \exp\left(-\frac{1}{2} \frac{(y-y_0)^2}{W_L}\right) \right)^{\frac{1}{2}} \quad (2)$$

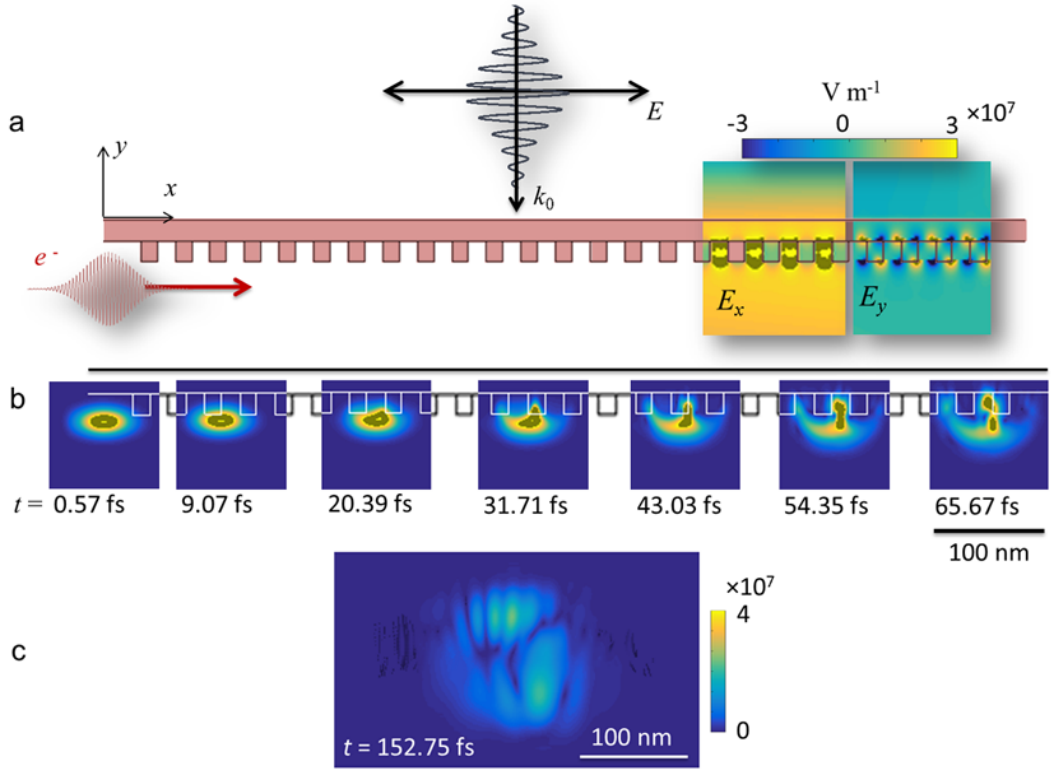

**Figure 2.** Inverse Smith–Purcell effect. (a) An electron pulse with longitudinal and transverse broadenings of 5 nm and 10 nm, respectively, travels at a distance of 4 nm and parallel to a silicon grating. The silicon grating has a period of 33.2 nm and is illuminated with a laser pulse at a carrier wavelength of 830 nm and a temporal broadening of 80 fs. The spatial distribution of the electric field components at a given time is depicted in the inset. Snapshots of the spatial distribution of the electron wavefunction (b) at times depicted on each frame during the interaction with the grating and (c) after the electron has left the grating.

where $W_t$ and $W_L$ are the transverse and longitudinal broadenings, $k_x^e = m_0 V_e / \hbar$ and $(x_0, y_0)$ is the initial position of the pulse centre at $t = 0$. The length of the grating is $960\,\mu\text{m}$ so that the time of flight of the electron through the acceleration medium becomes approximately 80 fs. Because the laser excitation is perpendicular to the electron trajectory, the synchronicity condition between the electron pulse and the grating near-field can be simplified as $\Lambda = m V_e \lambda_0 / c$, where by assuming $m = 1$, $\Lambda = 33.2\,\text{nm}$ is obtained. In this regard, the first diffraction order in the Floquet's expansion term will be considered as the synchronous mode.

The polarization of the laser illumination is along the grating axis. Such an illumination couples efficiently to the TE$_z$ mode of the grating. The spatial distribution of the total electric near-field is depicted in the inset of figure 2(a). Apparently, although only the $m = 1$ mode in the Floquet expansion set is desired, the dominant term is $m = 0$, as for all gratings for which the grating period is smaller than the wavelength of the incident light. On the other hand, the synchronicity condition $V_e / c = \Lambda / \lambda_0$ does not allow for utilization of diffraction gratings for which $\Lambda > \lambda_0$. This fact implies that, although acceleration is possible, it will come at the cost of severe chirping of the electron wave function. However, it will be shown later that this kind of chirping will be beneficial in electron bunching.

As an electron pulse with initial broadenings of $W_t = 5\,\text{nm}$ and $W_L = 10\,\text{nm}$ travels through the accelerating medium, it is strongly scattered by the grating. Snapshots of the spatial distribution of the electron wavefunction $|\psi(\vec{r}, t_i)|$ (figure 2(b)), show that interacting with several elements of the grating will cause the electron pulse to get bunched along the longitudinal direction. Moreover, because of the asymmetry of the grating configuration, the electron wave packet is deflected away transversely in the direction normal to the grating, a condition known as defocusing [82]. The shape of the electron wave after it has left the grating demonstrates a pronounced defocusing in the transverse direction and a severe chirping in the longitudinal direction (figure 2(c)).

The $E_y$ component of the electric field is localized at the edges of the grating elements with an evanescent distribution in the vacuum and correspondingly large wave vectors, which imposes severe scattering potentials for the electron wave. Moreover, the sign of the $E_y$ component and also the vector potential component $A_y$ are opposite at the adjacent edges, which causes a part of the wavefunction to be scattered to $\pm y$ directions. This can be better understood by comparing the momentum distributions of the electron wave as $\tilde{\psi}(k_x^e, k_y^e; t) = \frac{1}{2\pi} \int_{-\infty}^{+\infty} dx \int_{-\infty}^{+\infty} dy \, \exp(-ik_x^e x) \exp(-ik_y^e y) \psi(x, y; t)$ before and after its interaction with the grating, as shown in Figures 3a and 3b, respectively. It is apparent that the electron wave is mostly decelerated. Considering that all the transversely diffracted signals are detected, the probability of the electron to have the longitudinal momentum $k_x^e$ (or velocity $\hbar k_x^e / m_0$) is obtained as

$$\Gamma_L(k_x^e; t \to +\infty) = \int_{-\infty}^{+\infty} dk_y \, |\tilde{\psi}(k_x^e, k_y^e; t \to +\infty)|^2,$$ which is shown in figure 3(c).

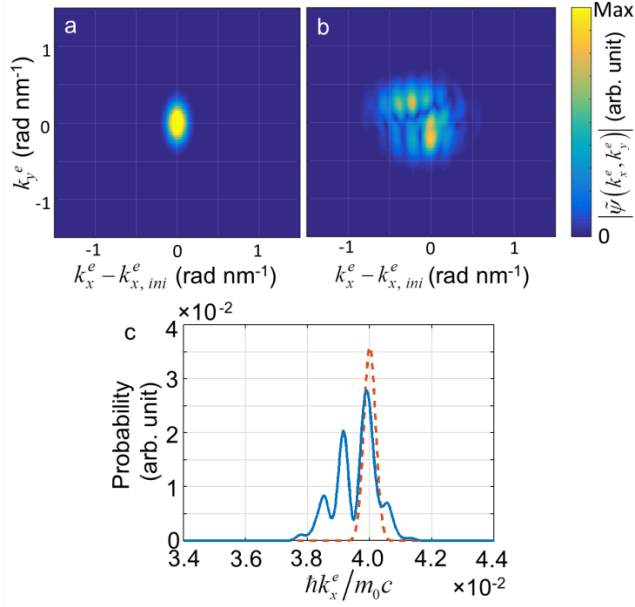

**Figure 3.** Electron wavefunction in momentum space. (a) Initial distribution of the Gaussian wavefunction and (b) the final distribution of the electron wave function before and after its interaction with the grating, respectively. (c) Probability distribution of the electron versus the normalized longitudinal component of the velocity $\hbar k_x^e / m_0 c$. $k_\alpha^e = p_\alpha / \hbar$ is the electron wavenumber and $p_\alpha$ is the momentum of the electron along the $\alpha$-axis. Dashed line: initial, and solid line: final probability distribution of the electron wave function.

Moreover, several resonances are observed in the momentum distribution of the electron wave function along the longitudinal momentum axis $k_x^e$, with an equidistant spacing of $\delta k_x^e = 2\pi / \Lambda$. Considering the phase matching condition as $\omega / V_e = 2\pi / \Lambda$, these resonances are due to the emissions of photons at the energy of $E_{ph} = \hbar V_e 2\pi / \Lambda$, which is exactly the laser carrier energy. Because the broadening of the electron provides sufficiently long time for multiple electron-photon interaction, even up to third-harmonic photon emission and absorption takes place, in a very similar way to multiple electron–photon scattering in PINEM [11, 45].

It has been discussed in Ref. [51], that a symmetrical force pattern should be employed to circumvent the defocusing of the electrons in interaction with the exponentially decaying near-field of the grating, at least on the axis of the accelerator. In order to examine this proposal numerically, a symmetric configuration is considered by placing two gratings parallel to each other and illuminating the whole structure with two mutually coherent phase-stabilized counter-propagating laser beams with the same specifications as mentioned above (see figure 4). In this way, only the even mode of the grating is excited, for which the $E_x$ field component has an even distribution, while both $E_y$ and $H_z$ field components have odd distributions along the y-axis. An initially Gaussian electron wave function at the carrier energy of 408 eV and with $W_t = 8$ nm and $W_L = 15$ nm travels through the grating along the symmetric axis. The gap between the gratings is considered as 30 nm.

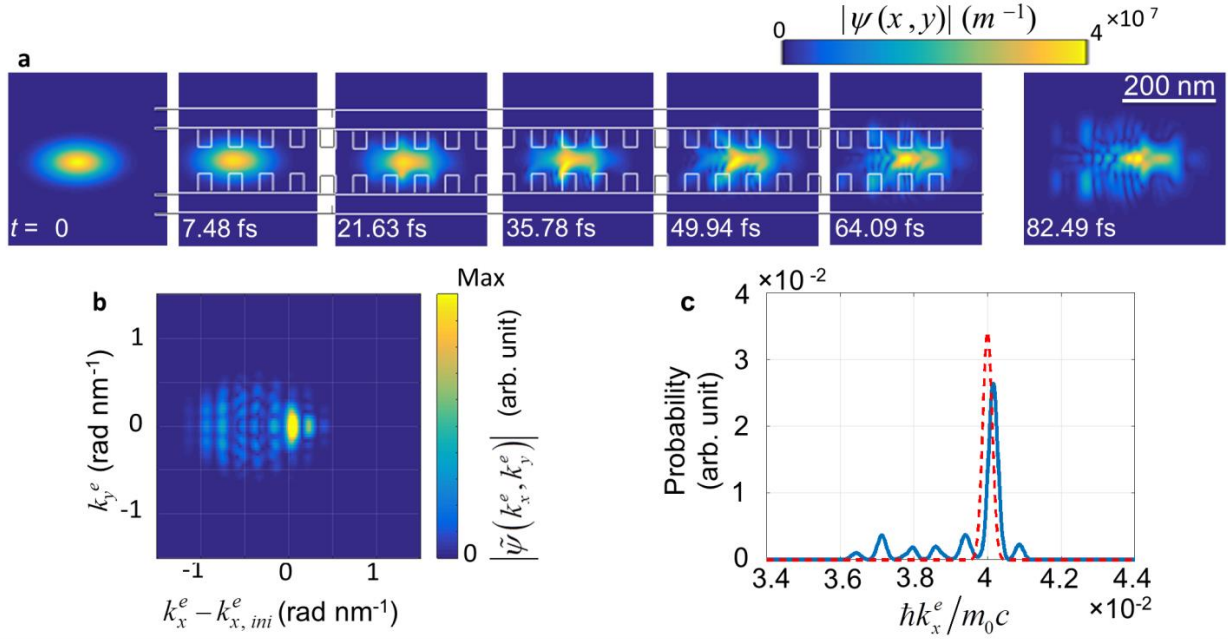

**Figure 4.** A Gaussian electron wavefunction interacting with a double grating illuminated with light in a symmetric phase-stabilized way. (a) Snapshots of the electron wavefunction at times depicted on each frame, (b) momentum representation of the electron wavefunction after interaction with the grating, and (c) probability distribution of the electron versus the normalized longitudinal component of the velocity $\hbar k_x^e / m_0 c$. $k_\alpha^e = p_\alpha / \hbar$ is the electron wavenumber and $p_\alpha$ is the momentum of the electron along the $\alpha$-axis. Dashed line: initial, and solid line: final probability distribution of the electron wave function.

It is anticipated that a symmetric force pattern enforces the momentum transferred from the light to the grating to be mostly along the longitudinal direction, so that the electron will be trapped in the transverse direction, quite similar to the photon trapping in quantum wells. In this way, even after few grating elements the electron wavefunction is getting bunched (figure 4(a)). The strong chirping of the electron in interaction with the grating is apparent from the final distribution after the electron has left the grating. Interestingly, the simulation results reveal that focusing happens only within a certain spatial region along the electron pulse, especially in the leading part of the wavefunction, whereas a strong defocusing is observed in the trailing part.

Indeed, the gap between the gratings supports a standing-wave pattern for the near-field components with large wave-vectors, which causes the electron to diffract transversely, in a very similar way to the Kapitza–Dirac effect. However, the Kapitza–Dirac effect is known to be a purely elastic scattering mechanism taking place with standing wave patterns in free space. In contrast, the diffractions which occur here are due to the interaction of the electrons with the near-field of the grating, and hence are happening in an inelastic way. Indeed, this inelastic diffraction phenomenon is another source of defocusing of the electron wave packet after the interaction with the gratings. Representation of the electron wavefunction in momentum space demonstrates that multiple loss and gain peaks occur along the longitudinal axis, whereas several diffraction orders are also observed in momentum space (figure 4(b)).

Computing $\Gamma_L\left(k_x^e;t\to+\infty\right)$ reveals that this configuration indeed leads to a net acceleration of the electron wave packet, though an acceleration gradient of only 4 MeV m$^{-1}$ is concluded. For computing the acceleration gradient, the rate of increase in the kinematic energy of the center of the pulse is considered as $m_0 V_e\, dV_e/dx$. Several energy-loss peaks are observed, which prohibit a net acceleration of the electron wave packet. In fact, because the initial electron energy is low, its acceleration takes place quite fast, and after few periods the synchronicity condition is not satisfied anymore. This effect is called dephasing [52]. The grating length after which dephasing happens is approximated as only 280 nm. Therefore it is not surprising that the electron cannot reach the highest possible acceleration gradient. In order to circumvent this difficulty, the phase velocity of the acceleration mode of the grating should remain synchronous with the electron. One solution is to introduce a multistage accelerator, by incorporating many tapered gratings of different periodicities [65]. Another solution is to introduce a chirped grating, as considered here.

The chirped double grating is designed in such a way as to maintain the synchronicity condition with the electron wave function by increasing the period in an adiabatic way. We start with $\Lambda_i = V_e \lambda_0/c$ and aim at reaching $\Lambda_f = 2 V_e \lambda_0/c$ within 800 nm (figure 5(a)). The spatiotemporal behaviour of the electron wavefunction during the interaction by such a grating is shown in figure 5(a). The leading as well as the trailing parts of the wave function are defocused, but the centre of the pulse is focused. This is somehow expected, as the synchronized longitudinal forces resulting from the x-component of the vector potential are always concomitant with the transverse components which are out of phase with the longitudinal component. The overall longitudinal dispersion and electron-pulse expansion, however, is much less than the case of non-chirped symmetrical grating (figure 4). As a consequence, the final longitudinal broadening of the electron pulse is also less. The momentum representation of the electron pulse as shown in figure 5(b) again reveals several energy loss and gain peaks. However, in contrast to the previous cases, energy-gain peaks are more pronounced, which is due to the acceleration of the centre of the electron wave-packet in a perfectly synchronized way through the whole interaction length (see figure 5(c)). As a consequence, an acceleration gradient of 7.5 MeV m$^{-1}$ is observed.

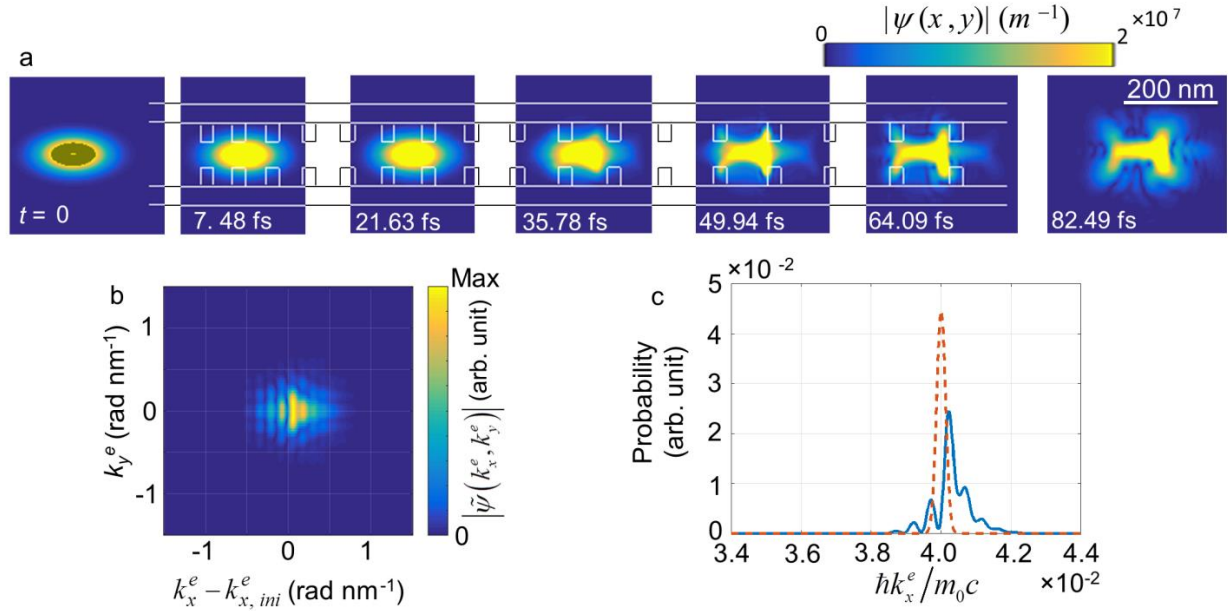

**Figure 5.** A Gaussian electron wavefunction interacting with a double chirped grating illuminated with light in a symmetric phase-stabilized way. (a) Snapshots of the electron wavefunction at times depicted on each frame, (b) momentum representation of the electron wavefunction after interaction with the grating, and (c) probability distribution of the electron versus the normalized longitudinal component of the velocity $\hbar k_x^e/m_0 c$. $k_\alpha^e = p_\alpha/\hbar$ is the electron wavenumber and $p_\alpha$ is the momentum of the electron along the $\alpha$-axis. Dashed line: initial, and solid line: final probability distribution of the electron wave function.

An interesting aspect of such numerical treatments is the ability to examine the contribution of each individual part of the Hamiltonian to the spatiotemporal behaviour of the electrons in the electromagnetic fields. For a relativistic electron pulse (initial kinetic energies higher than 50 keV), it is well justified to approximate the interaction Hamiltonian by the only leading term $\hat{H}_{int} = -(\hbar q/m_0)\vec{A}\cdot\vec{k}^e$, as has been considered in developing the theory of PINEM [11, 44]. However, for slow electrons, this assumption may lead to severe misunderstandings. First of all, the interaction Hamiltonian stated above leads to energy-loss and energy-gain peaks symmetrically distributed over the loss and gain parts of the spectrum. It is apparent from above-mentioned results that this is not the case here. To understand the reason behind this controversy, it is sufficient to assume a slowly varying electron wave packet as $\psi(x,y;t) = \psi_0(x,y;t)\exp(i\vec{k}^e\cdot\vec{r} - i\Omega t)$, where $\hbar\Omega = 0.5 m_0 V_e^2$ and $\hbar\vec{k}^e = \hat{x} m_0 V_e$ are the carrier energy and momentum of the initial electron wave packet. Substituting this wave packet into eq. (1), will lead to $(-\hbar^2/2m_0)\nabla^2\psi_0 - (i\hbar e/m_0)\vec{A}\cdot\vec{\nabla}\psi_0 - (\hbar^2/m_0)i\vec{k}^e\cdot\vec{\nabla}\psi_0 + (\hbar e/m_0)\vec{A}\cdot\vec{k}^e\psi_0 + (q^2/2m_0)|\vec{A}|^2\psi_0 + q\varphi\psi_0 = i\hbar\partial\psi_0/\partial t$, which is the equation of motion for the slowly varying term $\psi_0$. Apparently, the terms $-\hbar^2\nabla^2\psi_0/2m_0$ and $-i\hbar^2\vec{k}\cdot\vec{\nabla}\psi_0/m_0$ are responsible for the dispersion, chirping, and expansions of an initially Gaussian wave packet in free space, without considering any electromagnetic interaction. The term $\hbar e\vec{A}\cdot\vec{k}^e\psi_0/m_0 = \hbar e A_x k_x^e \psi_0/m_0$ causes the acceleration, as well as the symmetrically distributed energy loss and gain peaks [11]. It is the $-i\hbar e\vec{A}\cdot\vec{\nabla}\psi_0/m_0$ term which is not time-reversely symmetric and is the main cause of bunching, inelastic transverse diffraction, as well as asymmetric energy loss and gain peaks.

To better understand the chirping effect, the simulations here are to be compared with the case of a Gaussian electron wave packet propagating in free space [83]. By inserting the initial wave packet shown in eq. (2) in a free space propagator, we will find that (see Appendix)

$$\psi(x,y;t) = \frac{1}{\sqrt{2\pi}} \frac{\sqrt{W_L}}{\sqrt{W_L^2 + i\hbar t/2m_0}} \frac{\sqrt{W_t}}{\sqrt{W_t^2 + i\hbar t/2m_0}} \exp(ik_x^e x - i\omega_e t)$$
$$\times \exp\left(-\frac{(x - \hbar k_x^e t/m_0)^2}{4(W_L^2 + i\hbar t/2m_0)}\right) \exp\left(-\frac{y^2}{4(W_t^2 + i\hbar t/2m_0)}\right) \quad (3)$$

By a close analysis of eq. (3), it is revealed that an initially Gaussian wave packet experiences a chirp as well as expansion. The time dependent phase of this wave packet versus time and space is given by

$$\varphi(r,t) = ik_x^e x - i\omega_e t \left[ \left( \hbar t x / 2m_0 k_x^e + W_L^4 - \left( x/2k_x^e \right)^2 \right) \Big/ W_L^4 + \left( \hbar t / 2m_0 \right)^2 \right] + i \left[ \hbar y^2 / 2m_0 \left( W_t^4 + \left( \hbar t / 2m_0 \right)^2 \right) \right] t.$$

In other words, the center of the pulse will propagate with the constant phase of $\varphi_0 = \omega_e t$, which will be understood by setting $x = \hbar k_x^e t / m_0$ and $y = 0$ in $\varphi(x, y, t)$. However, it is not in general correct for other parts of the wave packet. In order to avoid significant changes in the energy spectrum, we should have $t \ll 2m_0 W_{L,t}^2 / \hbar$, which means that for $W_L = 15$ nm, $t \ll 394$ fs should be satisfied. Moreover, at sufficiently long times a significant decrease of the carrier energy should be observed. Considering now the interaction of a Gaussian wavepacket with the optical grating, the electron is accelerated instead of experiencing a pure bunching, due to the chirping of the Gaussian wave packet in already free space. Interestingly, the time threshold $t_s = 2m_0 W_{L,t}^2 / \hbar$ is related to the square of longitudinal broadening, which means that, for a sufficiently long pulse, the change in the carrier energy should be not pronounced, but the electron wave packet experiences bunching, as we will see in the following (see figures 6 and 7).

It is already apparent from the results presented above that coherent electron wave packets do interact in a different way with optical gratings in comparison with monolithic classical electrons at an equivalent centre of mass energy. Specifically, distinct diffraction orders observed along the transverse direction are a pure consequence of the wave behaviour of the electrons, very much like an electromagnetic wave traversing a diffraction grating. Moreover, multiple energy-loss and energy-gain peaks, which form due to multiple electron–photon scatterings, are not observed in a classical particle-in-cell numerical method.

Considering the facts stated above, it would be interesting to investigate the interaction of electron pulses at different longitudinal broadenings with the chirped-grating DLA introduced above. It is anticipated that more localized electron wave packets, achieved by making $W_{t,L}$ smaller, can be better accelerated provided that the near synchronicity condition is satisfied. On the other hand, vacuum is highly dispersive for non-relativistic electrons, and if no lenses or collimation system is incorporated [84] a well-localized electron pulse expands rapidly by propagating at distances as short as few hundreds of nanometers. In this regard, a compromise between the acceptable spatial broadening of the pulse and the best achievable acceleration gradient is to be searched for. In all the cases considered below, the carrier envelope phase is tuned in such a way that the $E_x$ field component experiences a maximum at the time when the electron reaches the first grating element, whereas the delay of the optical excitation is controlled to manifest best synchronization with the electron. In other words, the centre of the electron wave packet and the centre of the optical pulse concomitantly reach the middle of the grating. Moreover, the broadening of the optical pulse is considered to be equal to the travelling time of the centre of the electron wave packet along the grating (78 fs). This case however is an optimal condition for acceleration, if the dispersion effect could be ignored. In practice, there is no such control on the entrance phase. Furthermore, getting off-synchronism may be advantageous for acceleration, at least statistically even if we ignore dispersion.

Our investigations show that, for wave packets with $W_t < 4$ nm, the expansion in the transverse direction is quite severe during the propagation process. We choose to maintain the transverse broadening of the initial wave packet as $W_t = 8$ nm and only discuss the effect of the longitudinal broadening on the final shape of the electron wave function (figure 6). The broadenings considered here are just trial tests. An interesting consequence is that the larger the electron-wave packet broadening the better is the bunching effect. This spatiotemporal analysis reveals that electron bunching occurs when the electron wave packet is broad enough to cover several elements of the gratings. As this is the case for electrons emitted from state-of-the-art electron guns, either field emission or photoemission, one can expect that multiple scattering which happens within one single electron wave packet interacting with several elements of the gratings causes the electron wave packet to become bunched. Moreover, the shape of an electron wave packet with only $W_L = 5$ nm (time broadening $W_s = W_L/V_e = 417$ as) is much better preserved and the wavefunction is also not bunched. At the leading part of the wavefunction defocusing still occurs. The implication of the spatiotemporal chirping behaviour in momentum space unravels the occurrence of diffraction orders in the transverse direction, regardless of the electron-wave packet longitudinal broadening (figure 7). Because the angular distribution of the diffraction is relevant to the electron velocity, as shown in figure 7, the effect is not a pure elastic effect. However, still for the centre of the wave packet, the scattering can be considered elastic. According to the Kapitza-Dirac effect, an electron scattered from a standing wave light, undergoes a two-photon-scattering process, as a result a net transverse-momentum transfer equal to $P^e = 2n\hbar k^{ph}$ will be observed, where $k^{ph}$ is the photon wavenumber and $n$ is an integer. Considering this, the electron will be diffracted by an associated wavenumber of $k_y^e = 2nk^{ph}$. Considering $\lambda^{ph} = 30$ nm, which is equal to the gap between the gratings, $k_y^e = n\, 0.419\, \text{rad}\, \text{nm}^{-1}$ will be obtained, which agrees perfectly well with the numerical results obtained. It should be pointed out here that in contrast with the usual Kapitza-Dirac which is a result of the interaction of a plane wave electron with a free-space standing wave, the effect considered here is a pure near-field effect, happening because of the enhanced $E_y$ component of the electric field excited at the edges of the grating elements (see Figure 1a). The near-field distribution of the grating provides significantly large momentum to facilitate strong electron–photon coupling; a consequence of this is the clear diffraction orders happening here even by incorporating a moderate light intensity, which is not easily addressable in the Kapitza–Dirac effect. However, still the momentum transfer here is a two–photon process, quite similar to the Kapitza-Dirac effect.

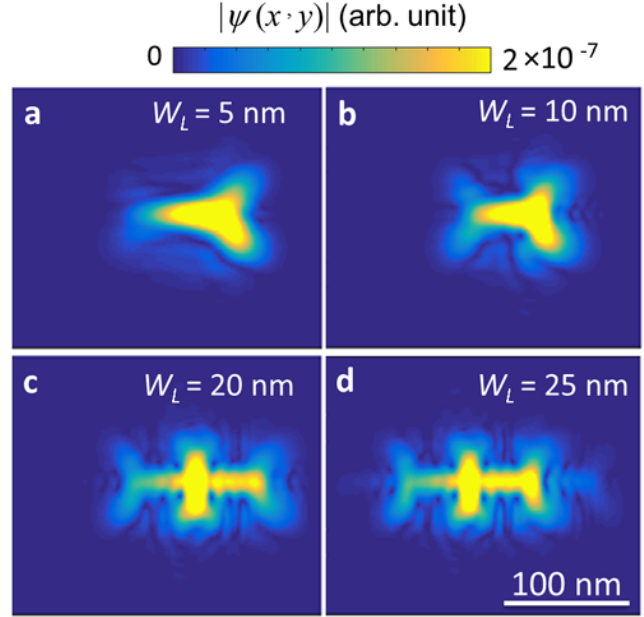

**Figure 6.** The spatial distribution of the absolute of the electron wavefunction, after the electron wave packet has left the grating, captured at $t = 82.5$ fs, for initial longitudinal electron broadenings of (a) $W_L = 5$ nm, (b) $W_L = 10$ nm, (c) $W_L = 20$ nm, and (d) $W_L = 25$ nm.

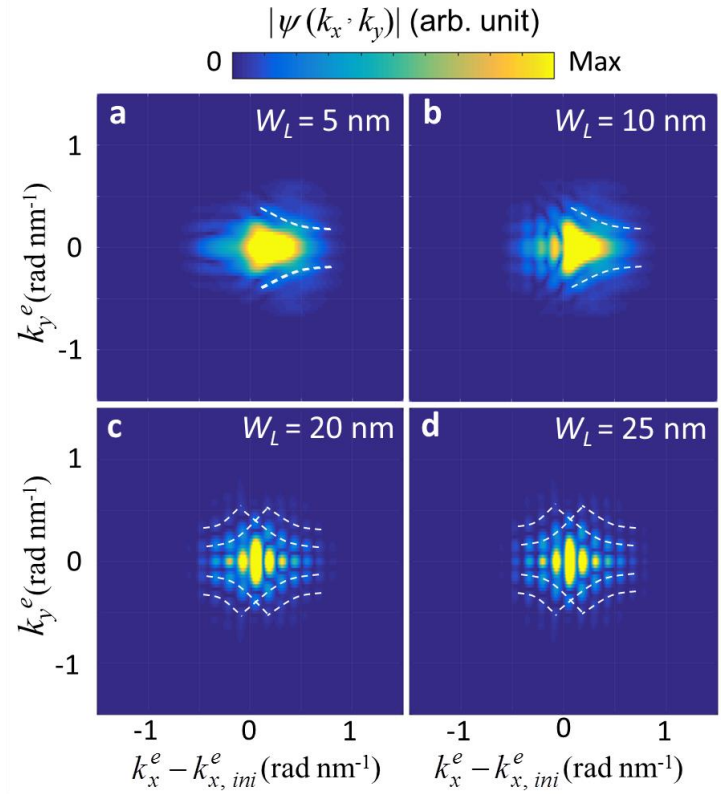

**Figure 7.** Absolute of the electron wavefunction in momentum space, after interaction with the grating, for initial longitudinal electron broadenings of (a) $W_L = 5\,\text{nm}$, (b) $W_L = 10\,\text{nm}$, (c) $W_L = 20\,\text{nm}$, and (d) $W_L = 25\,\text{nm}$.

The signature of the spatiotemporal bunching in energy–momentum space is the appearance of distinct resonances along the longitudinal momentum axis, causing electron energy loss and energy-gain peaks (figure 7c and d). A quantum mechanical interpretation of similar effects has been deliberately developed within the context of PINEM experiments [11], for which an electron wave packet interacts with the near-field of a sample, and loses quanta of photon energies. However, a clear distinction between PINEM and the electron-photon scatterings happening in the inverse Smith–Purcell effect is a net acceleration of the centre of the wave packet, due to the satisfaction of the synchronicity (or phase-matching [85]) condition. This net acceleration causes the resonance peaks to not appear symmetrically at the gain and loss regions.

An interesting result of this investigation is that the more localized the electron wave packet in the longitudinal direction, the better it is accelerated (figure 8). This is because that the multiple electron photon scatterings does not happen for short electron wave packets, due to the short interaction time of the single electron wave packet with light, which as a consequence does not allow the quantum mechanical interferences to take place. In other words, the light energy will be preserved for a net acceleration of the whole electron wave packet, instead of being transferred to nonlinear and multiple scattering processes.

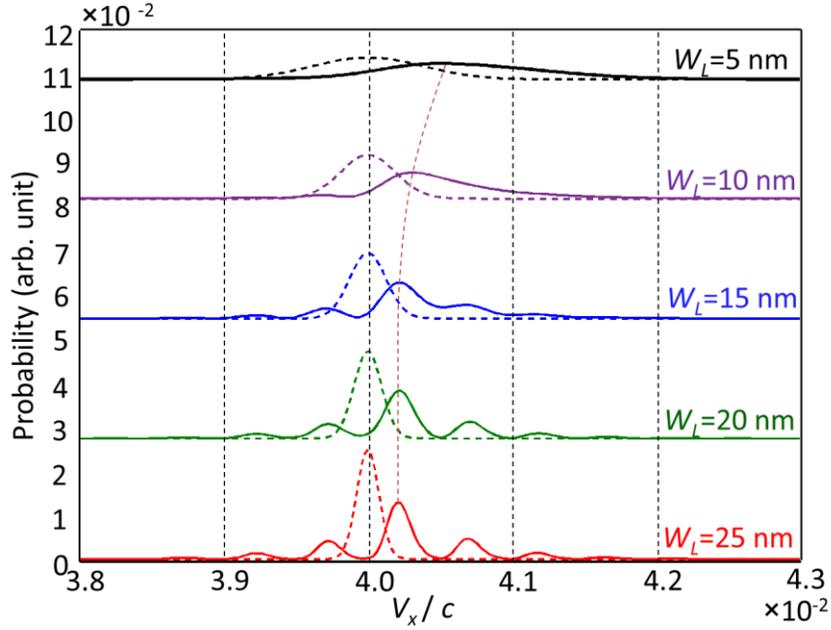

**Figure 8.** Probability distribution of the electron versus the normalized longitudinal component of the velocity $\hbar k_x^e / m_0 c$, taken for various initial longitudinal broadenings as depicted in the figure. Dashed line: initial, and solid line: final probability distribution of the electron wave function.

### 4. Conclusion

A self-consistent Maxwell-Schrödinger numerical toolbox is developed which helps to investigate the quantum aspects of the electron-photon interactions, beyond the adiabatic assumptions and undepleted pump approximation. Such a toolbox will have many applications in both understanding the experiments with matter waves, and also in exploring the possibilities to shape the matter waves [86]. This toolbox is used to study the interaction of single electron pulses with lights and gratings, a phenomenon referred to as inverse Smith-Purcell effect, which is incorporated regularly in DLAs. Interesting consequences of such an investigation is the bunching of single electron wavepackets in a fully coherent way, which only happens for electron pulses of sufficient longitudinal broadening, in such a way that they can cover few grating elements. A fully quantum mechanical aspect of the interaction is the occurrence of several diffraction peaks in the transverse direction. Although it has been proposed that a symmetric grating may be incorporated to make the electron wave packet focused, the Kapitza–Dirac effect can violate the assumptions made using the Lorentzian mechanics.

Electrons at low energies are easily recoiled and bunched in interaction with gratings and light, which might not be desirable in applications at which a Gaussian shape should be preserved. However, these consequences might be employed for an efficient electron wave packet shaping, such as bunched electron ways to be incorporated in coherent emission [87], and even making the electron wave packet chiral, as will be discussed in a future contribution.

## Acknowledgements

I sincerely thank Prof. P. A. van Aken and Dr. W. Sigle for fruitful discussions on the topic. The scholarship from the Max Planck Gesellschaft is greatly acknowledged.

## Appendix: Propagation of a Gaussian electron wave packet in free space

The propagator for an arbitrary wave packet propagating in a 2-dimensional free space can be obtained as

$$\psi(x,y;t) = \frac{1}{(2\pi)^2} \iint dk_x dk_y \, \tilde{\psi}(k_x, k_y; t=0) e^{-i\frac{\hbar}{2m_0}(k_x^2+k_y^2)t} e^{ik_x x + ik_y y} \tag{A1}$$

where $\tilde{\psi}(k_x, k_y; t=0)$ is the two-dimensional Fourier transform of the electron wave function at time $t = 0$, defined as $\tilde{\psi}(k_x, k_y; t=0) = \iint dx\,dy\, \psi(x,y;t=0) \exp(-ik_x x - ik_y y)$. Considering the initial wave function of eq. (2), $\tilde{\psi}(k_x, k_y; t=0) = \sqrt{8\pi W_L W_t} \exp\left(-W_L^2 (k_x - k_x^e)^2\right) \exp\left(-W_t^2 k_y^2\right)$ will be obtained. Inserting this in eq. (A1), the wave function at an arbitrary time and space will be obtained (see eq. (3)).

Equation (3) can be divided into 3 terms as $\psi(x,y;t) = F_1(x,y;t) F_2(x,y;t) F_3(x,y;t)$ where $F_1$ is the term containing the energy chirp and phase of the wave packet as

$$\begin{aligned} F_1(x,y;t) = & \exp(ik_x^e x - i\omega_e t) \\ & \times \exp\left( i\omega_e t \frac{(x/k_x^e)^2 + (\hbar t/m_0)^2 - 2\hbar t x/m_0 k_x^e}{4\left(W_L^4 + (\hbar t/2m_0)^2\right)} \right) \\ & \times \exp\left( + \frac{i\hbar t y^2/2m_0}{4\left(W_t^4 + (\hbar t/2m_0)^2\right)} \right) \end{aligned} \tag{A2}$$

Eq. (A2) explains the chirping effect and dispersion by propagation of the wave function in time and space. The expansion of the wave packet in time can be described by the second term as

$$F_2(x,y;t) = \exp\left( -\frac{W_L^2 (x - \hbar k_x^e t/m_0)^2}{4\left(W_L^4 + (\hbar t/2m_0)^2\right)} \right) \cdot \exp\left( -\frac{W_t^2 (y)^2}{4\left(W_t^4 + (\hbar t/2m_0)^2\right)} \right) \tag{A3}$$

and the third term is the renormalization factor due to the propagation in time, given by

$$F_3(x,y;t) = \frac{1}{\sqrt{2\pi}} \frac{\sqrt{W_L}}{\sqrt{W_L^2 + i\hbar t/2m_0}} \frac{\sqrt{W_t}}{\sqrt{W_t^2 + i\hbar t/2m_0}} \tag{A4}$$

## References

[1] Larmor J 1897 LXIII. On the theory of the magnetic influence on spectra; and on the radiation from moving ions *Philosophical Magazine Series 5* **44** 503-12
[2] Ginzburg V L 1993 Radiation by Uniformly Moving Sources - Vavilov-Cherenkov Effect, Doppler-Effect in a Medium, Transition Radiation and Associated Phenomena *Prog. Opt.* **32** 269-312
[3] Avdonina N B and Pratt R H 1999 Bremsstrahlung spectra from atoms and ions at low relativistic energies *J. Phys. B: At., Mol. Opt. Phys.* **32** 4261-76
[4] Smith S J and Purcell E M 1953 Visible Light from Localized Surface Charges Moving across a Grating *Phys. Rev.* **92** 1069
[5] Pantell R H, Soncini G and Puthoff H E 1968 Stimulated Photon-Electron Scattering *IEEE J. Quantum Electron.***Qe 4** 905
[6] Aamodt K, et al, 2008 The ALICE experiment at the CERN LHC *J. Instrum.* **3** S08002
[7] Gover A and Sprangle P 1981 A Unified Theory of Magnetic Bremsstrahlung, Electrostatic Bremsstrahlung, Compton-Raman Scattering, and Cerenkov-Smith-Purcell Free-Electron Lasers *IEEE J. Quantum Electron.* **17** 1196-215
[8] Adamo G, Ou J Y, So J K, Jenkins S D, De Angelis F, MacDonald K F, Di Fabrizio E, Ruostekoski J and Zheludev N I 2012 Electron-Beam-Driven Collective-Mode Metamaterial Light Source *Phys. Rev. Lett.* **109**, 217401
[9] Ackermann W, et al, 2007 Operation of a free-electron laser from the extreme ultraviolet to the water window *Nat. Photonics* **1** 336-42
[10] de Abajo F J G 2010 Optical excitations in electron microscopy *Rev. Mod. Phys.* **82** 209-75
[11] Park S T, Lin M M and Zewail A H 2010 Photon-induced near-field electron microscopy (PINEM): theoretical and experimental *New J. Phys.* **12**, 123028
[12] Barwick B, Flannigan D J and Zewail A H 2009 Photon-induced near-field electron microscopy *Nature* **462** 902-6
[13] Feist A, Echternkamp K E, Schauss J, Yalunin S V, Schafer S and Ropers C 2015 Quantum coherent optical phase modulation in an ultrafast transmission electron microscope *Nature* **521** 200
[14] Piazza L, Lummen T T A, Quinonez E, Murooka Y, Reed B W, Barwick B and Carbone F 2015 Simultaneous observation of the quantization and the interference pattern of a plasmonic near-field *Nat. Commun.* **6**
[15] Kroll N M 1982 Sensitivity of a Laser Driven Grating Linac to Grating Errors *Aip Conf. Proc.* 237-43
[16] Esarey E, Schroeder C B and Leemans W P 2009 Physics of laser-driven plasma-based electron accelerators *Rev. Mod. Phys.* **81** 1229-85
[17] Shimoda K 1962 Proposal for an Electron Accelerator Using an Optical Maser *Appl. Opt.* **1** 33-5
[18] England R J, et al, 2014 Dielectric laser accelerators *Rev. Mod. Phys.* **86** 1337-89
[19] Peralta E A, Soong K, England R J, Colby E R, Wu Z, Montazeri B, McGuinness C, McNeur J, Leedle K J, Walz D, Sozer E B, Cowan B, Schwartz B, Travish G and Byer R L 2013 Demonstration of electron acceleration in a laser-driven dielectric microstructure *Nature* **503** 91
[20] Kealhofer C, Schneider W, Ehberger D, Ryabov A, Krausz F and Baum P 2016 All-optical control and metrology of electron pulses *Science* **352** 429-33
[21] Kirchner F O, Gliserin A, Krausz F and Baum P 2014 Laser streaking of free electrons at 25 keV *Nat. Photonics* **8** 52-7
[22] Garciamolina R, Grasmarti A, Howie A and Ritchie R H 1985 Retardation Effects in the Interaction of Charged-Particle Beams with Bounded Condensed Media *J. Phys. C: Solid State Phys.* **18** 5335-45
[23] Talebi N, Ozsoy-Keskinbora C, Benia H M, Kern K, Koch C T and van Aken P A 2016 Wedge Dyakonov Waves and Dyakonov Plasmons in Topological Insulator Bi2Se3 Probed by Electron Beams *ACS Nano* (to be published)
[24] de Abajo F J G, Rivacoba A, Zabala N and Yamamoto N 2004 Boundary effects in Cherenkov radiation *Phys. Rev. B* **69**

[25]     Talebi N, Sigle W, Vogelgesang R, Esmann M, Becker S F, Lienau C and van Aken P A 2015 Excitation of Mesoscopic Plasmonic Tapers by Relativistic Electrons: Phase Matching versus Eigenmode Resonances *ACS Nano* **9** 7641-8
[26]     Groß P, Esmann M, Becker S F, Vogelsang J, Talebi N and Lienau C 2016 Plasmonic nanofocusing – grey holes for light *Advances in Physics: X* **1** 297-330
[27]     Talebi N, Ogut B, Sigle W, Vogelgesang R and van Aken P A 2014 On the symmetry and topology of plasmonic eigenmodes in heptamer and hexamer nanocavities *Appl. Phys. A: Mater. Sci. Process.* **116** 947-54
[28]     Kociak M and Stephan O 2014 Mapping plasmons at the nanometer scale in an electron microscope *Chem. Soc. Rev.* **43** 3865-83
[29]     Bosman M, Keast V J, Watanabe M, Maaroof A I and Cortie M B 2007 Mapping surface plasmons at the nanometre scale with an electron beam *Nanotechnology* **18**
[30]     Goris B, Guzzinati G, Fernandez-Lopez C, Perez-Juste J, Liz-Marzan L M, Trugler A, Hohenester U, Verbeeck J, Bals S and Van Tendeloo G 2014 Plasmon Mapping in Au@Ag Nanocube Assemblies *J. Phys. Chem. C* **118** 15356-62
[31]     Gu L, Sigle W, Koch C T, Ogut B, van Aken P A, Talebi N, Vogelgesang R, Mu J L, Wen X G and Mao J 2011 Resonant wedge-plasmon modes in single-crystalline gold nanoplatelets *Phys. Rev. B* **83** 195433
[32]     Ogut B, Talebi N, Vogelgesang R, Sigle W and van Aken P A 2012 Toroidal Plasmonic Eigenmodes in Oligomer Nanocavities for the Visible *Nano Lett.* **12** 5239-44
[33]     Talebi N, Sigle W, Vogelgesang R, Koch C T, Fernandez-Lopez C, Liz-Marzan L M, Ogut B, Rohm M and van Aken P A 2012 Breaking the Mode Degeneracy of Surface Plasmon Resonances in a Triangular System *Langmuir* **28** 8867-73
[34]     Talebi N 2014 A directional, ultrafast and integrated few-photon source utilizing the interaction of electron beams and plasmonic nanoantennas *New J. Phys.* **16** 053021
[35]     Coenen T and Polman A 2014 Optical Properties of Single Plasmonic Holes Probed with Local Electron Beam Excitation *ACS Nano* **8** 7350-8
[36]     Luo C, Ibanescu M, Johnson S G and Joannopoulos J D 2003 Cerenkov radiation in photonic crystals *Science* **299** 368-71
[37]     Le Thomas N, Alexander D T L, Cantoni M, Sigle W, Houdre R and Hebert C 2013 Imaging of high-Q cavity optical modes by electron energy-loss microscopy *Phys. Rev. B* **87** 155314
[38]     Yamamoto N, de Abajo F J G and Myroshnychenko V 2015 Interference of surface plasmons and Smith-Purcell emission probed by angle-resolved cathodoluminescence spectroscopy *Phys. Rev. B* **91** 125144
[39]     Kesar A S 2010 Smith-Purcell radiation from a charge moving above a grating of finite length and width *Phys. Rev. Spec. Top.--Accel. Beams* **13** 022804
[40]     Mizuno K, Pae J, Nozokido T and Furuya K 1987 Experimental-Evidence of the Inverse Smith-Purcell Effect *Nature* **328** 45-7
[41]     de Abajo F J G and Kociak M 2008 Electron energy-gain spectroscopy *New J Phys* **10**
[42]     Asenjo-Garcia A and de Abajo F J G 2013 Plasmon electron energy-gain spectroscopy *New J. Phys.* **15** 103021
[43]     Talebi N, Sigle W, Vogelgesang R and van Aken P 2013 Numerical simulations of interference effects in photon-assisted electron energy-loss spectroscopy *New J. Phys.* **15** 053013
[44]     Howie A 2011 Photon interactions for electron microscopy applications *Eur. Phys. J.: Appl. Phys.* **54** 33502
[45]     de Abajo F J G, Asenjo-Garcia A and Kociak M 2010 Multiphoton Absorption and Emission by Interaction of Swift Electrons with Evanescent Light Fields *Nano Lett.* **10** 1859-63
[46]     Fallahi A and Kartner F 2014 Field-based DGTD/PIC technique for general and stable simulation of interaction between light and electron bunches J. Phys. B: At., Mol. Opt. Phys. **47** 234015
[47]     Verboncoeur J P 2005 Particle simulation of plasmas: review and advances *Plasma Phys. Controlled Fusion* **47** A231

[48]   Vay J L 2008 Simulation of beams or plasmas crossing at relativistic velocity *Phys. Plasmas* **15** 056701
[49]   Smirnova O, Spanner M and Ivanov M 2008 Analytical solutions for strong field-driven atomic and molecular one- and two-electron continua and applications to strong-field problems *Phys. Rev. A* **77** 033407
[50]   Wolkow D M 1935 On a mass of solutions of the Dirac equation. *Zeitschrift für Physik* **94** 250-60
[51]   Naranjo B, Valloni A, Putterman S and Rosenzweig J B 2012 Stable Charged-Particle Acceleration and Focusing in a Laser Accelerator Using Spatial Harmonics, *Phys. Rev. Lett.* **109** 164803
[52]   Breuer J, McNeur J and Hommelhoff P 2014 Dielectric laser acceleration of electrons in the vicinity of single and double grating structures-theory and simulations J. Phys. B: At. Mol. Opt. Phys. **47** 234004
[53]   Kapitza P L and Dirac P A M 1933 The reflection of electrons from standing light waves. *Proc. Cambridge Philos. Soc.* **29** 297-300
[54]   Batelaan H 2007 Colloquium: Illuminating the Kapitza-Dirac effect with electron matter optics *Rev. Mod. Phys.* **79** 929-41
[55]   Batelaan H 2000 The Kapitza-Dirac effect *Contemp. Phys.* **41** 369-81
[56]   Talezer H and Kosloff R 1984 An Accurate and Efficient Scheme for Propagating the Time-Dependent Schrodinger-Equation *J. Chem. Phys.* **81** 3967-71
[57]   Talebi N, Shahabadi M, Khunsin W and Vogelgesang R 2012 Plasmonic grating as a nonlinear converter-coupler *Opt. Express* **20** 1392-405
[58]   Talebi N and Shahabadi M 2010 All-optical wavelength converter based on a heterogeneously integrated GaP on a silicon-on-insulator waveguide *J. Opt. Soc. Am. B* **27** 2273-8
[59]   Askar A and Cakmak A S 1978 Explicit Integration Method for Time-Dependent Schrodinger Equation for Collision Problems *J. Chem. Phys.* **68** 2794-8
[60]   Busch K, Konig M and Niegemann J 2011 Discontinuous Galerkin methods in nanophotonics *Laser Photonics Rev.* **5** 773-809
[61]   Kosloff R 1988 Time-Dependent Quantum-Mechanical Methods for Molecular-Dynamics *J. Phys. Chem. -Us* **92** 2087-100
[62]   Cerjan C, Kosloff D, Kosloff R and Reshef M 1985 A Nonreflecting Boundary-Condition for Discrete Acoustic and Elastic Wave-Equations *Geophysics* **50** 705-8
[63]   Talebi N 2016 Optical modes in slab waveguides with magnetoelectric effect *J. Optics-Uk* **18**
[64]   Ferrario M, et al, 2014 IRIDE: Interdisciplinary research infrastructure based on dual electron linacs and lasers *Nucl. Instrum. Methods Phys. Res., Sect. A* **740** 138-46
[65]   Breuer J and Hommelhoff P 2013 Laser-Based Acceleration of Nonrelativistic Electrons at a Dielectric Structure *Phys. Rev. Lett.* **111** 134803
[66]   Gross P, Piglosiewicz B, Schmidt S, Park D J, Vogelsang J, Robin J, Manzoni C, Farinello P, Cerullo G and Lienau C 2015 Controlling the Motion of Strong-Field, Few-Cycle Photoemitted Electrons in the Near-Field of a Sharp Metal Tip *Springer Proceedings in Physics, Ultrfast Phenomenon XIX* **162** 659-62
[67]   Kasmi L, Kreier D, Bradler M, Riedle E and Baum P 2015 Femtosecond single-electron pulses generated by two-photon photoemission close to the work function *New J. Phys.* **17** 033008
[68]   Hoffrogge J, Stein J P, Kruger M, Forster M, Hammer J, Ehberger D, Baum P and Hommelhoff P 2014 Tip-based source of femtosecond electron pulses at 30 keV *J. Appl. Phys. (Melville, NY, U. S.)* **115** 094506
[69]   Piglosiewicz B, Schmidt S, Park D J, Vogelsang J, Gross P, Manzoni C, Farinello P, Cerullo G and Lienau C 2014 Carrier-envelope phase effects on the strong-field photoemission of electrons from metallic nanostructures *Nat. Photonics* **8** 38-43
[70]   Aidelsburger M, Kirchner F O, Krausz F and Baum P 2010 Single-electron pulses for ultrafast diffraction *Proc. Natl. Acad. Sci. U. S. A.* **107** 19714-9
[71]   Kruger M, Schenk M, Forster M and Hommelhoff P 2012 Attosecond physics in photoemission from a metal nanotip *J. Phys. B: At., Mol. Opt. Phys.* **45** 074006

[72]     Herink G, Solli D R, Gulde M and Ropers C 2012 Field-driven photoemission from nanostructures quenches the quiver motion *Nature* **483** 190-3
[73]     Bionta M R, Chalopin B, Champeaux J P, Faure S, Masseboeuf A, Moretto-Capelle P and Chatel B 2014 Laser-induced electron emission from a tungsten nanotip: identifying above threshold photoemission using energy-resolved laser power dependencies *J. Mod. Opt.* **61** 833-8
[74]     Schroder B, Sivis M, Bormann R, Schafer S and Ropers C 2015 An ultrafast nanotip electron gun triggered by grating-coupled surface plasmons *Appl. Phys. Lett.* **107** 231105
[75]     Muller M, Kravtsov V, Paarmann A, Raschke M B and Ernstorfer R 2016 Nanofocused Plasmon-Driven Sub-10 fs Electron Point Source *ACS Photonics* **3** 611-9
[76]     Echternkamp K E, Herink G, Yalunin S V, Rademann K, Schafer S and Ropers C 2016 Strong-field photoemission in nanotip near-fields: from quiver to sub-cycle electron dynamics *Appl. Phys. B: Lasers Opt.* **122** 80
[77]     Baum P 2013 On the physics of ultrashort single-electron pulses for time-resolved microscopy and diffraction *Chem. Phys.* **423** 55-61
[78]     Muller M, Paarmann A and Ernstorfer R 2014 Femtosecond electrons probing currents and atomic structure in nanomaterials *Nat. Commun.* **5** 5292
[79]     Bainbridge A R, Myers C W B and Bryan W A 2016 Femtosecond few- to single-electron point-projection microscopy for nanoscale dynamic imaging *Struct. Dyn.-Us* **3** 023612
[80]     Vogelsang J, Robin J, Nagy B J, Dombi P, Rosenkranz D, Schiek M, Gross P and Lienau C 2015 Ultrafast Electron Emission from a Sharp Metal Nanotaper Driven by Adiabatic Nanofocusing of Surface Plasmons *Nano Lett.* **15** 4685-91
[81]     Gliserin A, Walbran M and Baum P 2016 A high-resolution time-of-flight energy analyzer for femtosecond electron pulses at 30 keV *Rev. Sci. Instrum.* **87** 033302
[82]     Naranjo B, Valloni A, Putterman S and Rosenzweig J B 2012 Stable Charged-Particle Acceleration and Focusing in a Laser Accelerator Using Spatial Harmonics *Phys. Rev. Lett.* **109** 164803
[83]     Friedman A, Gover A, Kurizki G, Ruschin S and Yariv 1988 A Spontaneous and stimulated emission from quasi free electrons *Rev. Mod. Phys.* **60** 471-535
[84]     Gliserin A, Apolonski A, Krausz F and Baum P 2012 Compression of single-electron pulses with a microwave cavity *New J. Phys.* **14** 073055
[85]     Arbel M, Abramovich A., Eichenbaum A. L., Gover A., Kleinman H., Pinhasi Y., Yakover I. M., 2001 Super-radiant and stimulated super-radiant emission in a pre-bunched beam free electron maser *Phys. Rev. Lett.* **86** 2561-64
[86]     Verbeeck J, Guzzinati G, Clark L, Juchtmans R, Van Boxem R, Tian H, Beche A, Lubk A and Van Tendeloo G 2014 Shaping electron beams for the generation of innovative measurements in the (S)TEM *C. R. Phys.* **15** 190-9
[87]     Arbel M, Eichenbaum A L, Pinhasi Y, Lurie Y, Tecimer M, Abramovich A, Kleinman H, Yakover I M and Gover A 2000 Super-radiance in a prebunched beam free electron maser *Nucl. Instrum. Methods Phys. Res., Sect. A* **445** 247-52